\documentclass[12pt]{article}
\usepackage[esperanto,english]{babel}
\usepackage{epsfig}
\usepackage{parallel} 
\newcommand{\ppunu}[1]{}  \newcommand{\ppdu}[1]{#1}
\newcommand{\comentario}[1]{}
\ppdu{
\newlength{\pplw}\setlength{\pplw}{0.48\textwidth}
\newlength{\pprw}\setlength{\pprw}{0.50\textwidth}

\newcommand{\ppp}{\ParallelPar}
\newcommand{\ppn}{\noindent}              %p/ nao fazer parag
\newcommand{\ppl}[1]{\ParallelLText{\selectlanguage{esperanto}#1}}
\newcommand{\ppr}[1]{\ParallelRText{\selectlanguage{english}#1}\ppp}
\newcommand{\ppln}[1]
{\ParallelLText{\ppn \selectlanguage{esperanto}#1}} %p/ nao parag
\newcommand{\pprn}[1]
{\ParallelRText{\ppn \selectlanguage{english}#1}\ppp} %p/ nao parag

\newcommand{\ppR}[1]{\ParallelRText{#1}}

\newcommand{\ppsection}[3][0ex]{\vspace{2em} 
\ppl{\section{#2} \vspace{#1}} \ppa \nopagebreak
\ppR{\section{#3}} \ppp \nopagebreak}

\newcommand{\bea}{\vspace{-5mm}\begin{eqnarray}}
\newcommand{\eea}{\end{eqnarray}}
\newcommand{\dd}{\mathrm d}
}

\ppunu{
\newcommand{\ppl}[1]{\selectlanguage{esperanto}#1}
\newcommand{\ppln}[1]{\noindent \selectlanguage{esperanto}#1}
\newcommand{\ppr}[1]{\selectlanguage{english}}
\newcommand{\pprn}[1]{\selectlanguage{english}}
\newcommand{\ppsection}[3][0ex]{\section{#2}}

\newcommand{\bea}{\vspace-2ex}{\begin{eqnarray}}
\newcommand{\eea}{\end{eqnarray}}
}

\title{Oscillation of a rigid rod in the special relativity \\ Oscilado de rigida stango \^ce la speciala relativeco}
\author{F.M. Paiva \\ 
{\small Departamento de F\'\i sica, Unidade Humait\'a II, Col\'egio Pedro II} \\
{\small Rua Humait\'a 80, 22261-040  Rio de Janeiro-RJ, Brasil; fmpaiva@cbpf.br} 
\vspace{.7ex} \\
A.F.F. Teixeira \\
{\small Centro Brasileiro de Pesquisas F\'\i sicas} \\
{\small 22290-180 Rio de Janeiro-RJ, Brasil; teixeira@cbpf.br}}
\date{}%\date{\today}%{5--a de Oktobro de 2011}

\begin{document}
\selectlanguage{english}
\maketitle

\begin{abstract}\selectlanguage{english}
In the special relativity, a rigid rod slides upon itself, with one extremity oscillating harmonically. We discovered restrictions in the amplitude of the motion and in the length of the rod, essential to eliminate unphysical solutions.  

\ppdu{\selectlanguage{esperanto}
\^Ce la speciala relativeco, rigida stango movi\^gas sur si mem, kun unu fino oscilante harmonie. Ni malkovris limigajn kondi\^cojn pri la amplitudo de movado kaj pri la longo de stango, necesegaj por elimini ne-fizikajn solvojn.}
\end{abstract}
\selectlanguage{english}

\ppdu{
\begin{Parallel}[v]{\pplw}{\pprw}
%\begin{Parallel}[v]{}{}
}

\ppdu{\section*{\vspace{-2em}}\vspace{-2ex}}   %PORQUE PRECISO DISTO ?

\ppsection[0.6ex]{Enkonduko}{Introduction}

\ppln{Rigida stango komence ripozas, etendita sur akso $x$ de inercia sistemo de referenco $S_0$\,, inter pozicioj $x_a = 0$ kaj $x_b = L > 0$\,. En momento $t = 0$ la fino $a$ komencas movadon $x_a(t_a)$ sur akso $x$\,. Ni nomas $[t_a,x_a]$ la postaj eventoj de $a$ en sistemo $S_0$\,.}
\pprn{A rigid rod is initially at rest, laying along the $x$-axis of an inertial system of reference $S_0$\,, between the positions $x_a = 0$ and $x_b = L > 0$\,. In the moment $t = 0$ the extremity $a$ starts a motion $x_a(t_a)$ on the $x$-axis\,. We call $[t_a,x_a]$ the succeeding events of $a$ in the system $S_0$\,.}

\ppl{La fino $b$ rigide akompanas la movadon de $a$\,, anka\u u sur la akso. Tio estas, $b$ konservas la distancon $L$ al $a$ en la sinsekvaj inerciaj sis\-temoj de referenco $S_v$ kie $a$  momente restas. Vidu~\cite{Born}, \cite[pa\^go\,289]{Moller}, \cite[pa\^go\,50]{Rindler}, \cite{GiannoniGron} --  \cite{Lrt2}, pro detaloj pri rigida movado en speciala relativeco. Nomante $[t_b,x_b]$ la eventoj de $b$ en sistemo $S_0$\,, Nikoli\'c~\cite{Nikolic} montris, ke en iu ajn movado $x_a(t_a)$ okazas}
\ppr{The extremity $b$ follows rigidly the motion of $a$\,, also on the axis. That is, $b$ maintains the distance $L$ to $a$ in the successive inertial reference systems $S_v$ where $a$ is momentarily at rest. See \cite{Born}, \cite[page\,289]{Moller}, \cite[page\,50]{Rindler}, \cite{GiannoniGron} -- \cite{Lrt2}), for details on the rigid motion in the special relativity. Calling $[t_b,x_b]$ the events of $b$ in the system $S_0$\,, Nikoli\'c~\cite{Nikolic} showed that for any motion $x_a(t_a)$ one has}

\bea                                                                                \label{uai}%01
x_b=x_a+L\gamma\,, \hspace{3mm} t_b=t_a+\frac{L}{c^2}v\gamma\,; \hspace{3mm}  v:=v_a(t_a)=\frac{\dd x_a}{\dd t_a}\,, \hspace{3mm} \gamma:=(1-v^2/c^2)^{-1/2}\,. 
\eea

\ppln{Simpla kalkulo~\cite[Sek.\,9]{Lrt1} montras, ke $\dd x_b/\dd t_b$, en momento $t_b$, egalas $\dd x_a/\dd t_a$ en momento $t_a$\,. Konsekvence, \^ciuj punktoj de la rigida stango restas en la sinsekvaj inerciaj sistemoj $S_v$\,.}
\pprn{A simple calculation~\cite[Sec.\,9]{Lrt1} shows that $\dd x_b/\dd t_b$\,, in moment $t_b$, is equal to $\dd x_a/\dd t_a$ in moment $t_a$\,. Consequently, all points of the rigid rod are at rest in the successive inertial systems $S_v$\,.}

\ppl{Ni studas la okazon kun fino $a$ de stango movi\^gante harmonie en sistemo $S_0$\,. Se $A$ estas la duonamplitudo de movado, kaj $\omega$ estas la frekvenco, tial}
\ppr{We study the case with the extremity $a$ of the rod moving harmonically in the system $S_0$\,. If $A$ is the half-amplitude of the motion, and $\omega$ is the frequency, then}

\bea                                                                              \label{comcw1}%02
x_a=A(1-\cos\omega t_a)\,. 
\eea 

\ppl{Por tiu movado de $a$\,, la ne-harmonia movado de $b$ estas esprimita per la parametro $t_a$ kiel}
\ppr{For that motion of $a$\,, the non-harmonic motion of $b$ is expressed via the parameter $t_a$ as} 

\bea                                                                              \label{comcw2}%03
x_b=x_a+\frac{L}{\sqrt{1-(A\omega/c)^2\sin^2\omega t_a}}\,, \hspace{3mm} t_b=t_a+\frac{(AL\omega/c^2)\sin\omega t_a}{\sqrt{1-(A\omega/c)^2\sin^2\omega t_a}}\,.
\eea

\ppln{Sen perdi generalecon, ni konsideru $\omega=1$ kaj $c=1$ (vidu~\cite{volta}). Tiu simpligas ekvaciojn~(\ref{comcw1}) kaj (\ref{comcw2}) al}
\pprn{Without loss of generality, we consider $\omega=1$ and $c=1$ (see~\cite{volta}). This simplifies equations (\ref{comcw1}) and (\ref{comcw2}) to}

\bea                                                                               \label{semcw}%04
x_a=A(1-\cos t_a)\,,\; x_b=x_a+\frac{L}{\sqrt{1-A^2\sin^2t_a}}\,,\; t_b=t_a+\frac{AL\sin t_a}{\sqrt{1-A^2\sin^2t_a}}\,.
\eea 

\ppl{Tiel, la movado de $b$ estas karakterizita per paro $[A\,; L\,]$. La duonamplitudo $A$ estas lim\-igita al $A<1$\,, por eviti ke la rapido $A\sin t_a$ de $a$ atingu valoron $c$. \^Car $v_b(t_b)=v_a(t_a)$\,, tial anka\u u la rapido de $b$ neniam estos $c$\,. Ekzemple, la paro $[0,\!5\,; 1,\!8\,]$ generas movadojn por $a$ kaj $b$ kiel figuro~\ref{tresfig}a montras.}
\ppr{Thus the motion of $b$ is characterized by a pair $[A\,; L\,]$. The semiamplitude $A$ is bound to $A<1$, to prevent velocity $A\sin t_a$ of $a$ reaching the value $c$. Since $v_b(t_b)=v_a(t_a)$\,, then similarly the velocity of $b$ will never be $c$\,. For example, the pair $[0,\!5\,; 1,\!8\,]$ generates motions for $a$ and $b$ as figure~\ref{tresfig}a shows.}

\begin{figure}[t]                                                                         %Figura 1
\centerline{\epsfig{file=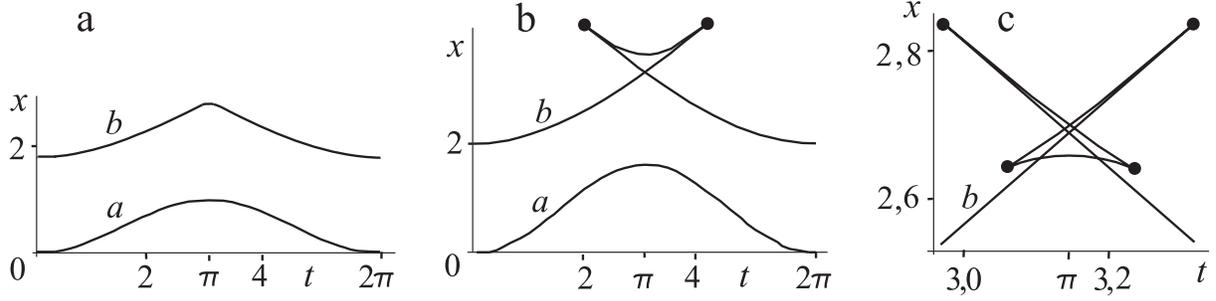,width=160mm}} %width=103mm,height=57mm                
\selectlanguage{esperanto}\caption{
Horizonta akso indikas la tempon $0<t<2\pi$ en inercia sistemo $S_0$, kaj la vertikala akso montras la serion de pozicioj $x_a$ (harmonia movado) kaj $x_b$ (ne-harmonia movado). La ne-konstanta distanco inter $b$ kaj $a$ en sistemo $S_0$ estas mezurita vertikale. La klino de la trajektorioj evidentigas, ke \^ciu rapido estas plieta ol $c$\,. Nigra disketo ($\bullet$) indikas eventon $[t_b,x_b]$ kie nefinia propra akcelo okazas. 
\newline {\bf a} La paro de parametroj $[A\,; L\,]=[0,\!5\,; 1,\!8\,]$ generas movadojn fizike eblajn. 
\newline {\bf b} La paro $[0,\!8\,; 2,\!0\,]$ permesas $b$ sammomente esti en e\^c 3 malsimilaj lokoj, en tempa intervalo $2,\!0<t<4,\!3$\,. Do tiu movado estas ne-fizika.  
\newline {\bf c} La paro $[0,\!88\,; 0,\!90\,]$ permesas $b$ sammomente esti en e\^c 5 malsimilaj lokoj.  Do tiu movado estas ne-fizika.
\vspace{2mm} 
\ppdu{\newline \selectlanguage{english}Figure \ref{tresfig}:
The horizontal axis indicates the time $0<t<2\pi$ of inertial system $S_0$, and the vertical axis shows the successive positions $x_a$ (harmonic motion) and $x_b$ (non-harmonic motion). The variable distance between $b$ and $a$ in the system $S_0$ is measured vertically. The slope of the trajectories makes evident that all velocities are shorter than $c$\,. A black disk ($\bullet$) indicates an event $[t_b,x_b]$ where an infinite proper acceleration occurs. 
\newline {\bf a} The pair of parameters $[A\,; L\,]=[0,\!5\,; 1,\!8\,]$ generates motions physically possible. 
\newline {\bf b} The pair $[0,\!8\,; 2,\!0\,]$ allows $b$ be present in even 3 different places in a same moment, in the time interval $2,\!0<t<4,\!3$\,. Thus this motion is unphysical. 
\newline {\bf c} The pair $[0,\!88\,; 0,\!90\,]$ allows $b$ be present in even 5 different places, simultaneously. Thus this motion is unphysical.}}
\label{tresfig}
\end{figure}

\ppl{Tamen, la nura kondi\^co $A<1$ ne tute eliminas ne-fizikajn solvojn de (\ref{semcw}b) kaj (\ref{semcw}c). Ekzemple, la paro $[A\,; L\,]=[0,\!8\,; 2,\!0\,]$ permesas finon $b$ esti en e\^c 3 malsimilaj pozicioj $x_b$ en sama momento $t_b$, kiel figuro~\ref{tresfig}b montras. Kaj la paro $[0,\!88\,; 0,\!90\,]$ permesas $b$ esti samtempe en e\^c 5 malsimilaj lokoj, kiel figuro~\ref{tresfig}c montras.}
\ppr{However, the restriction $A<1$ alone does not completely eliminate unphysical solutions of (\ref{semcw}b) and (\ref{semcw}c). For example, the pair $[A\,; L\,]=[0,\!8\,; 2,\!0\,]$ permits the extremity $b$ be present in even 3 different positions $x_b$ in a same moment $t_b$, as figure~\ref{tresfig}b shows. And the pair $[0,\!88\,; 0,\!90\,]$ permits $b$ be simultaneously in even 5 different places, as figure~\ref{tresfig}c shows.}

\ppl{\^Car tiuj matematikaj rezultoj estas fizike neakcepteblaj, tial ni devas trovi pluajn limigojn pri $A$ kaj $L$\,. Ni profitas la evidentajn abruptajn malordojn en figuroj~\ref{tresfig}b kaj \ref{tresfig}c, indikitaj per nigraj disketoj. Tiuj malordoj rezultas el nefiniaj akceloj, kiuj estas fizike neeblaj.}
\ppr{Since these mathematical results are physically unacceptable, we ought finding other limitations in $A$ and $L$\,. We profit the evident abrupt disorders in the figures \ref{tresfig}b and \ref{tresfig}c, indicated by small black disks. These disorders result from infinite accelerations, which are physically impossible.}

\ppsection{Propra akcelo}{Proper acceleration}

\ppln{Propra akcelo de punkto estas tiu akcelo mezurita en inercia sistemo de referenco kie la punkto momente restas. Vidu \cite{Born}, \cite[pa\^go\,22]{LL}, \cite[pa\^go\,13]{Moller}, \cite[pa\^go\,49]{Rindler}, \cite{GiannoniGron}, \cite{Nikolic}, por detaloj. Se la akcelo kaj la rapido de punkto estas paralelaj, tial la propra akcelo estas~\cite{Lrt1}}
\pprn{The proper acceleration of a point is the one measured in an inertial system where the point is momentarily at rest. See \cite{Born}, \cite[page\,22]{LL}, \cite[page\,13]{Moller}, \cite[page\,49]{Rindler}, \cite{GiannoniGron}, \cite{Nikolic}, for details. If the acceleration and the velocity of the point are parallel, then the proper acceleration is~\cite{Lrt1}}  

\bea                                                                              \label{proakc}%05
g(t):=\frac{\dd}{\dd t}\left(\frac{\dd x/\dd t}{\sqrt{1-(\dd x/\dd t)^2}}\right)\,.
\eea

\ppln{La fino $a$\,, kies movado $x_a(t_a)$ estas elektita en (\ref{semcw}a), havas propran akcelon}
\pprn{The extremity $a$, whose motion $x_a(t_a)$ was chosen in (\ref{semcw}a), has proper acceleration}

\bea                                                                                  \label{ga}%06
g_a(t_a)=\frac{A\cos t_a}{(1-A^2\sin^2t_a)^{3/2}}\,; 
\eea

\ppln{ni vidas, ke la akcelo $g_a$ estas \^ciam finia.}
\pprn{we see that the acceleration $g_a$ is always finite.}

\ppl{Por koni la propran akcelon $g_b$ de $b$ ni uzas anka\u u~(\ref{proakc}) kaj rememoras, ke $v_b(t_b)=v_a(t_a)$\,. La kalkulo do da\u urigas tiel:}
\ppr{To know the proper acceleration $g_b$ of $b$ we use also~(\ref{proakc}), and remember that $v_b(t_b)=v_a(t_a)$\,. The calculus then proceeds thus:} 

\bea                                                                                 \label{gb1}%07
g_b(t_b)= \frac{\dd}{\dd t_b}\left(\frac{v_a(t_a)}{\sqrt{1-v_a^2(t_a)}}\right)= \left(\frac{\dd t_a}{\dd t_b}\right)\frac{\dd}{\dd t_a} \left(\frac{v_a(t_a)}{\sqrt{1-v_a^2(t_a)}}\right)=\left(\frac{\dd t_a}{\dd t_b}\right)g_a(t_a)=\frac{g_a(t_a)}{\dd t_b/\dd t_a}\,.
\eea

\ppln{\^Car $g_a$ estas \^ciam finia, tial $g_b$ estas nefinia nur en momentoj $t_b$ tiaj ke la nomanto $\dd t_b/\dd t_a$ en (\ref{gb1}) estas nula. Uzante~(\ref{semcw}c) ni vidas, ke la respondaj momentoj $t_a$ obeas}
\pprn{Since $g_a$ is always finite, $g_b$ is infinite only in moments $t_b$ such that the denominator $\dd t_b/\dd t_a$ in (\ref{gb1}) is null. Using (\ref{semcw}c), we see that the corresponding moments $t_a$ obey}

\bea                                                                                 \label{gb2}%08
AL\cos t_a+(1-A^2\sin^2t_a)^{3/2}=0\,.
\eea

\ppln{Tio implicas $\cos t_a < 0$\,, tial $\pi/2<t_a<3\pi/2$\,.}
\pprn{This implies $\cos t_a < 0$\,, so $\pi/2<t_a<3\pi/2$\,.}

\ppl{Ekzemple, la paro $[A=0,\!8\,; L=2,\!0\,]$ en (\ref{gb2}) oferas solvon $\cos t_a=-0,\!143$\,, ekvivalente $t_{a1}=1,\!7$ kaj $t_{a2}=4,\!6$\,; uzante (\ref{semcw}c), tiuj $t_a$ donas $t_{b1}=2,\!0$ kaj $t_{b2}=4,\!3$\,; kaj uzante (\ref{semcw}b) ili donas $x_{b1}=x_{b2}=4,\!2$\,. Fakte en figuro~\ref{tresfig}b ni vidas, ke la eventoj $[t_{b1}, x_{b1}\,]$ kaj $[t_{b2}, x_{b2}\,]$ markas finojn de duobleco de movadoj de $b$, en sistemo $S_0$\,.}
\ppr{For example, the pair $[A=0,\!8\,; L=2,\!0\,]$ in (\ref{gb2}) has solution $\cos t_a=-0,\!143$\,, or equivalently $t_{a1}=1,\!7$ and $t_{a2}=4,\!6$\,; using (\ref{semcw}c), these $t_a$ give $t_{b1}=2,\!0$ and $t_{b2}=4,\!3$\,; and using (\ref{semcw}b) they give  $x_{b1}=x_{b2}=4,\!2$\,. In fact, in figure~\ref{tresfig}b we see that the events $[t_{b1}, x_{b1}\,]$ and $[t_{b2}, x_{b2}\,]$ mark endpoints of duplicity of motions of $b$, in the system $S_0$\,.}

\ppsection{La fizikaj okazoj}{The physical cases}

\ppln{Ni volas elmontri, klare kaj precize, la parojn $[A\,; L\,]$ fizike eblaj. Por tio, ni komence desegnas grafika\^{\j}ojn de $\dd t_b/\dd t_a$ kiel funkcio de $t_a$ por pluraj valoroj de paro $[A\,; L\,]$, kiel en figuro~\ref{figTodos}. En tiu figuro ni vidas, ke nur subfiguroj 3, 6 kaj 7 havas $\dd t_b/\dd t_a$ neniam nula, do estas la nuraj fizike eblaj.}
\pprn{We want to exhibit, with clarity and precision, the pairs $[A\,; L\,]$ physically possible. To that end, we start drawing graphics of $\dd t_b/\dd t_a$ against $t_a$ for a large number of values of the pair $[A\,; L\,]$, as in figure~\ref{figTodos}. In that figure, we see that only the subfigures 3, 6 and 7 do not have zeros of $\dd t_b/\dd t_a$\,, so they are the sole physically possible.} 

\ppl{Se ni atentas la serion de subfiguroj $3\rightarrow2\rightarrow1$ ni vidas, ke $\dd t_b/\dd t_a$ unue estas nula en subfiguro 2, en $t_a=\pi$\,, kiam anka\u u $\dd^2t_b/\dd {t_a}^2=0$\,. Tiu lasta ekvacio, uzante $t_b(t_a)$ oferita en (\ref{semcw}c), ekvivalentas}
\ppr{If we regard the sequence of subfigures $3\rightarrow2\rightarrow1$ we see that one zero one $\dd t_b/\dd t_a$ first emerges in subfigure 2, in $t_a=\pi$\,, where also $\dd^2t_b/\dd {t_a}^2=0$\,. This last equation, using $t_b(t_a)$ given in (\ref{semcw}c), is equivalent to}

\bea 																																								 \label{seg}%09
(3A^2-1-2A^2\sin^2t_a)\sin t_a = 0\,.  
\eea

\ppl{Ekvacio (\ref{seg}) helpas koni la fizike eblajn okazojn. Portante al (\ref{gb2}) la solvon $\sin t_a=0$ de (\ref{seg}), ni ricevas la hiperbolon}
\ppr{Equation (\ref{seg}) helps knowing the cases physically possible. Bringing into (\ref{gb2}) the solution $\sin t_a=0$ of (\ref{seg}), we get the hyperbola}

\bea                                                                                 \label{AL1}%10
L=\frac{1}{A}\,, 
\eea 

\ppln{desegnita per kontinua linio en la centra grafika\^{\j}o de figuro~\ref{figTodos}. La paroj $[A\,; L]$ sur a\u u super tiu hiperbolo donas solvojn fizike neeblaj. Por esplori la parojn sub la hiperbolo, ni bezonas konsideri la alian solvon de (\ref{seg}).}
\pprn{drawn in full line in the central graphic of the figure~\ref{figTodos}. The pairs $[A\,; L]$ on or above that hyperbola give solutions physically impossible. To explore the pairs under the hyperbola, we need consider the other solution of (\ref{seg}).} 

\begin{figure}[t]                                                                         %Figura 2
\centerline{\epsfig{file=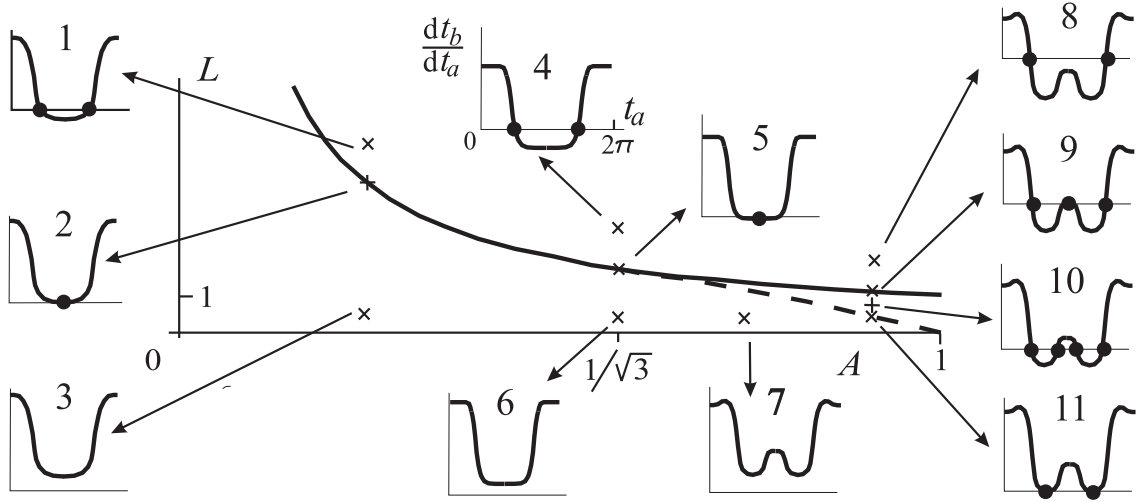,width=150mm}} %width=103mm,height=57mm                
\selectlanguage{esperanto}\caption{
En la centra grafika\^{\j}o estas hiperbolo $L=1/A$ per kontinua linio, kaj parabolo $L=(3\sqrt{3}/2)(1-A^2)$ se $A>1/\sqrt{3}$ per strekita linio. \^Cirka\u ue estas 11 skizoj de la kvalite malsimilaj grafika\^{\j}oj de $\dd t_b/\dd t_a$ kiel funkcio de $t_a$\,, dissemitaj en ebeno $(A\,; L\,)$. Nigra disketo ($\bullet$) indikas nefinian propran akcelon ($\dd t_b/\dd t_a=0$). Nur grafika\^{\j}oj 3, 6, kaj 7 estas fizike eblaj.  
\vspace{2mm}
\ppdu{\newline \selectlanguage{english}Figura \ref{figTodos}:
In the central graph are the hyperbola $L=1/A$ in full line, and the parabola $L=(3\sqrt{3}/2)(1-A^2)$ if $A>1/\sqrt{3}$ in dashed line. Around are 11 sketches of those graphs qualitatively differents of $\dd t_b/\dd t_a$ as function of $t_a$\,, distributed in in the plane $(A\,; L\,)$. A black disk ($\bullet$) indicates infinite proper acceleration ($\dd t_b/\dd t_a=0$). Only the graphs 3, 6, and 7 are physically possible.}}
\label{figTodos}
\end{figure}

\ppl{Tiu alia solvo estas $2A^2\sin^2t_a=3A^2-1$\,. \^Gi gravas se nur $A>1/\sqrt{3}$\,, celante ke $\sin t_a$ estu reela. Uzante tiun solvon en (\ref{gb2}), ni ricevas la arkon de parabolo}
\ppr{That other solution is $2A^2\sin^2t_a=3A^2-1$\,. It matters only if $A>1/\sqrt{3}$\,, so that $\sin t_a$ be real. Using that solution in (\ref{gb2}), we get the arc of parabola}

\bea                                                                                 \label{AL2}%11
L=\frac{3\sqrt{3}}{2}(1-A^2)\,, \hspace{2mm} 1/\sqrt{3}<A<1\,, 
\eea

\ppln{montrita per la strekita linio en la centra grafika\^{\j}o en figuro~\ref{figTodos}.}
\pprn{shown by the dashed line in the central graph of figure~\ref{figTodos}.}

\ppl{{\bf Resume}, la paroj $[A\,; L\,]$ fizike eblaj havas $L<1/A$ $\;$ se $\;$ $A<1/\sqrt{3}$, $\;$ kaj $\;$ havas $L<(3\sqrt{3}/2)(1-A^2)$ $\;$ se $\;$ $1/\sqrt{3}<A<1$\,.}
\ppr{{\bf In short}, the pairs $[A\,; L\,]$ physically possible have $L<1/A$ $\;$  if $\;$  $A<1/\sqrt{3}$\,, $\;$ and $\;$ have $L<(3\sqrt{3}/2)(1-A^2)$ $\;$ if $\;$ $1/\sqrt{3}<A<1$\,.}

\ppl{Ni konfirmas tiun konkludon atentante la serion $7\rightarrow11\rightarrow10\rightarrow9\rightarrow8$\,; tie ni vidas, ke $\dd t_b/\dd t_a$ ne nuli\^gas en subfiguro~7, kaj unue nuli\^gas en subfiguro~11, kie anka\u u  $\dd^2t_b/\dd{t_a}^2=0$ -- vidu la du nigrajn disketojn sur la strekata linio. Se ni pligrandigas $L$, \^ciu disketo tuj fari\^gas du, kiel en subfiguro~10. Pligrandigante $L$\,, la du internaj disketoj proksimi\^gas kaj fine unui\^gas kiel en subfiguro~9, kie denove $\dd^2t_b/\dd{t_a}^2=0$ en $t_a=\pi$\,. Fine, se $L$ estas malmulte pligranda, tiu centra disketo malaperas, kiel en subfiguro~8.}
\ppr{We confirm that conclusion in the sequence $7\rightarrow11\rightarrow10\rightarrow9\rightarrow8$\,; there we see that $\dd t_b/\dd t_a$ has no zero in the subfigure~7, and starts having zero in the subfigure~11, where also $\dd^2t_b/\dd{t_a}^2=0$ -- see the two black disks on the streched line. If we increase $L$\,, each disk soon becomes two disks, as in subfigure~10. Increasing $L$ a little more, the two disks approach and finally melt as in subfigure~9, where again $\dd^2t_b/\dd{t_a}^2=0$ in $t_a=\pi$\,. Eventually, if $L$ is a little larger, that central disk disappears, as in subfigure~8.} 

\ppl{Anka\u u la serio $6\rightarrow5\rightarrow4$ estas interesa. \^Gi okazas en vertikalo $A=1/\sqrt{3}$\,, kaj havas $\dd^3t_b/\dd{t_a}^3$ nulan se $t_a=\pi$\,. Do \^ciuj kurboj $\dd t_b/\dd t_a$ kun $A=1/\sqrt{3}$ havas nulan kurbecon en $t_a=\pi$\,, kiel ni vidas en figuro~\ref{figTodos}.}
\ppr{Also the sequence $6\rightarrow5\rightarrow4$ is interesting. It occurs in the vertical $A=1/\sqrt{3}$\,, and has the derivatives $\dd^3t_b/\dd{t_a}^3$ null in $t_a=\pi$\,. So all curves $\dd t_b/\dd t_a$ with $A=1/\sqrt{3}$ have null curvature in $t_a=\pi$\,, as we see in figure~\ref{figTodos}.}

\begin{figure}                                                                            %Figura 3
\centerline{\epsfig{file=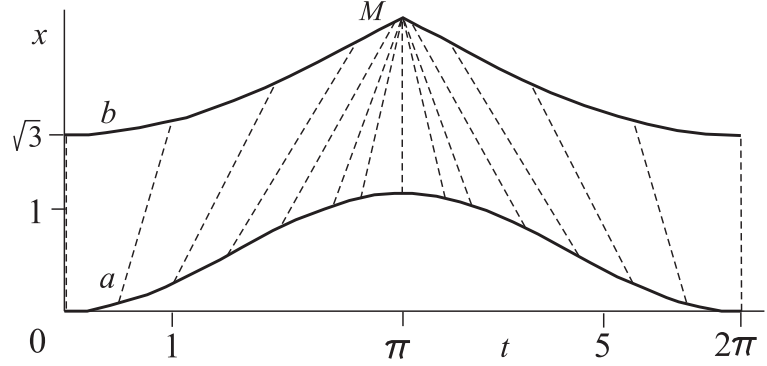,width=100mm}} %width=103mm,height=57mm                
\selectlanguage{esperanto}\caption{
Movadoj de $b$ kaj $a$ se $A=1/\sqrt{3}$ kaj $L=\sqrt{3}$. La strekitaj linioj kunigas parojn de samtempaj eventoj en la sinsekvaj inerciaj sistemoj de referenco kie $a$ kaj $b$ momente restas. Ni atentas amason de apartaj eventoj \^ce $a$ kiuj estas preska\u u samtempaj al evento $M=[\pi\,, 2A+L\,]$ \^ce $b$\,, en tiuj sistemoj. 
\vspace{2mm}
\ppdu{\newline \selectlanguage{english}Figure \ref{figSimul}:
Motions of $b$ and $a$ if $A=1/\sqrt{3}$ and $L=\sqrt{3}$. The dashed lines connect pairs of simultaneous events in the successive inertial reference systems where $a$ and $b$ are momentarily at rest. We remark a large quantity of separated events of $a$ that are almost simultaneous to the event $M=[\pi\,, 2A+L\,]$ of $b$\,, in these systems.}}
\label{figSimul}
\end{figure}

\ppsection{Komentoj}{Comments}

\ppln{Klarigo indas emfazon: la egaleco $v_b(t_b)=v_a(t_a)$ ne signifas, ke la rapidoj de $a$ kaj $b$ estas egalaj en sama momento de inercia sistemo $S_0$\,. Fakte, la egaleco $v_b(t)=v_a(t)$ okazas nur kiam $t_b=t_a$\,. Atentante (\ref{semcw}c), tio implicas $t_a=0, \pi, 2\pi, ...$ Nur en tiuj momentoj $t_a$\,, la rigida stango estas senmova en $S_0$\,, kaj elmontras sian propralongon $L$ en $S_0$\,.}
\pprn{A clarification deserves emphasis: the equality $v_b(t_b)=v_a(t_a)$ does not mean that the velocities of $a$ and $b$ are equal in a same moment in the inertial reference system $S_0$\,. In fact, the equality $v_b(t)=v_a(t)$ occurs only when $t_b=t_a$\,. According to (\ref{semcw}c), that implies $t_a=0, \pi, 2\pi, ...$ Only in these moments $t_a$\,, the rigid rod is immobile in $S_0$\,, and exhibits its properlength $L$ in $S_0$\,.}

\ppl{Subfiguro 5 de figuro~\ref{figTodos} evidentigas, ke la okazo $A=1/\sqrt{3}$ kaj $L=\sqrt{3}$ estas speciala. La plateco de kurbo $\dd t_b/\dd t_a$ en akso $t_a$ implicas amason de nulaj valoroj de $\dd t_b/\dd t_a$ \^cirka\u u $t_a=\pi$\,. Fakte, figuro~\ref{figSimul} montras, ke tiuokaze multaj apartaj eventoj de fino $a$ estas preska\u u samtempaj al evento $M:=[\pi\,; 2A+L\,]$ \^ce $b$\,, en la sinsekvaj inerciaj sistemoj de referenco kie amba\u u $a$ kaj $b$ restas.}
\ppr{Subfigure 5 of figure~\ref{figTodos} makes evident that the case $A=1/\sqrt{3}$ and $L=\sqrt{3}$ is special. The flatening of the curve $\dd t_b/\dd t_a$ on the axis $t_a$ implies a large number of zeros of $\dd t_b/\dd t_a$ near $t_a=\pi$\,. Indeed, figure~\ref{figSimul} shows that in this case many separated events of the extremity $a$ are almost simultaneous to the event $M:=[\pi\,; 2A+L\,]$ of $b$\,, in the successive inertial systems where both $a$ and $b$ are at rest.}

\ppl{Fine indas mencii, ke la unu sola punkto kun harmonia movado en la rigida stango povus esti interna, anstata\u u fina. Tiu \^ci artikolo estus facile adaptita por priskribi tiun pli \^generalan okazon.}
\ppr{Finally, it is worth mentioning that the sole point with harmonic motion in the rigir rod could be interior, instead of an extremity. This article would be easily adapted to describe that more general case.} 

\ppdu{\vspace{2em}
\ppl{\section*{~}\vspace{-1em}} \nopagebreak
\ppR{\section*{References}\vspace{-1em}} \ppp \nopagebreak
\vspace{-1.9em}}
\selectlanguage{esperanto}

\ppdu{\end{Parallel}}

\end{document}